\documentclass{elsart}

\usepackage{graphicx}

\usepackage{amssymb}

\begin{document}

\begin{frontmatter}



\title{Fine-grained Delaunay triangulation in a simulation of tumor spheroid growth}


\author[Trieste]{Alessio Del Fabbro}
\ead{delfabbro@ts.infn.it}
\author[Verona]{Roberto Chignola}
\ead{roberto.chignola@univr.it}
\author[Trieste]{Edoardo Milotti\corauthref{cor}}
\ead{milotti@ts.infn.it}
\corauth[cor]{Corresponding author}
\address[Trieste]{Dipartimento di Fisica, Universit\`a di Trieste and I.N.F.N. -- Sezione di Trieste \\ Via Valerio, 2 -- I-34127 Trieste, Italy }
\address[Verona]{Dipartimento Scientifico e Tecnologico, Facolt\`a di Scienze MM.FF.NN. \\ Universit\`a di Verona and I.N.F.N. -- Sezione di Trieste \\ Strada Le Grazie, 15 - CV1 -- I-37134 Verona, Italy}
\begin{abstract}
The simulation of many-particle systems often requires the detailed knowledge of proximity relations to reduce computational complexity and to provide a basis for specific calculations. Here we describe the basic scheme of a simulator of tumor spheroid growth: the calculation of mechanical interactions between cells and of the concentrations of diffusing chemicals requires a backbone provided by the Delaunay triangulation and the volumes of the associated Voronoi regions. Thus the Delaunay triangulation provides both the proximity relations needed to reduce the computational complexity and the basic structures that are needed to carry out the calculation of the biochemical interactions between cells and with the enviroment. A 3D version of the simulator uses the CGAL library as an essential component for the efficient computation of the Delaunay triangulation and of the Voronoi regions. 
\end{abstract}

\end{frontmatter}

\section{Introduction}
\label{intro}

A good understanding of the growth kinetics of tumors is essential to devise better and more effective therapeutic strategies \cite{norton}. Direct observation {\it in vivo} of the growth kinetics is not always possible, and a particularly useful {\it in vitro} technique uses multicell spheroids. Experimental multicell spheroids have volumes that range from about $10^{-4}$ mm$^{3}$ to about 1 mm$^3$ and contain as many as $10^6$ cells; they have a complexity which is intermediate between 2D cultures and tumors {\it in vivo} and display a growth kinetics that is very close to tumors {\it in vivo} \cite{sutherland,rob}.

Since there is no angiogenesis and thus no point transport of nutrients and oxygen, the local environment plays a very important role in the growth of tumor spheroids; experimental observations show that:
\begin{itemize}
\item spheroids are layered, and are characterized by an external layer of proliferating cells, by a buried layer of quiescent cells, and by a necrotic core, which is made up of cells either starved or asphyxiated;
\item there are strong, measurable oxygen and glucose gradients;
\item the structure is not fixed but behaves like a high-viscosity fluid, with a convective transport of cells from the outer layers to the core \cite{mp};
\item the shape is mostly ball-like, but some spheroids develop fractal-like structures, like dendritic structures on the surface or holes in the bulk.
\end{itemize}
Multicells tumor spheroids thus act as a clean experimental setup that reproduces many microscopic features of tumors {\it in vivo}, and captures most of the complex non-linear interactions among cells and with the environment.
Unfortunately, accurate measurements on tumor spheroids last at least a couple of months, and environmental conditions are neither very well controllable nor reproducible, and for this reason we are now developing a novel simulator of tumor spheroid growth. Our final aim is a full-fledged {\it in silico} simulator of the growth and proliferation of tumor cells, a sort of virtual laboratory where we may experiment at will and have access to all growth variables. Here we describe one important feature of this simulator, the inclusion of CGAL to compute the basic structures that we need to calculate the diffusion of nutrients, oxygen and other chemicals in the multicell cluster, and that are used to speed up the mechanical evolution of the cluster.

\section{Simulator structure and role of diffusion in cell clusters}
\label{struct}

\begin{figure}
\begin{center}
\includegraphics[width = 5in]{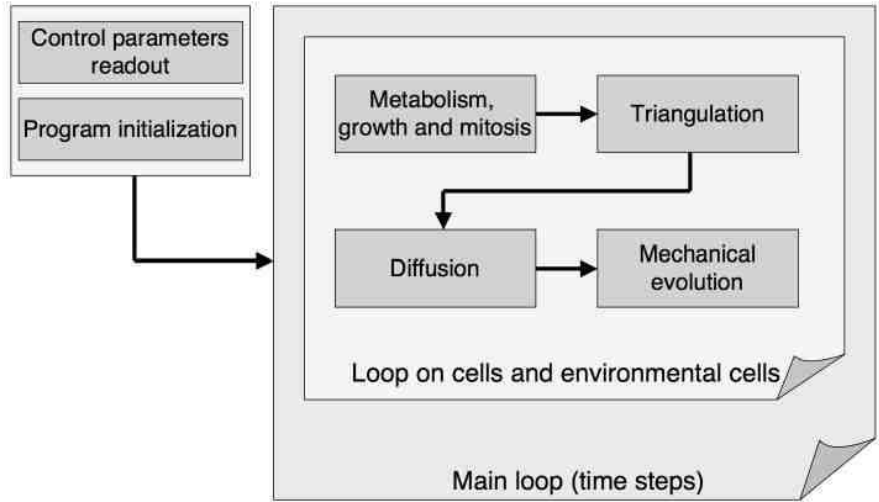}
\caption{Conceptual scheme of the simulator. An initialization step, which includes a definition of the parameters of the simulated experiment, is followed by a loop with fixed time steps. At each time step the cells consume nutrients and oxygen, grow and proliferate. Cells interact with neighboring cells, both mechanically and biochemically, and this step requires a detailed knowledge of the proximity relations which is provided by the Delaunay triangulation module. The Delaunay triangulation also provides the backbone for the calculation of Fickian diffusion of oxygen and other substances. Finally, a mechanical evolution step, computes the new positions of the cells after the internal stresses (due to growth) and the external forces have been accounted for; this step also needs the proximity relations to reduce the computational complexity.}
\label{Schematics}
\end{center}
\end{figure}

The simulator structure has already been described in \cite{cm} and is shown schematically in figure \ref{Schematics}. The main loop of the simulator includes the following steps: 
\begin{itemize}
\item Metabolic step. As they absorb nutrients from the environment, cells grow and proliferate: this step advances a kind of internal cell clock according to a rather detailed model of cell metabolism \cite{cm2}.
Cells also divide and generate new individuals: as a consequence this step redefines the cluster structure, and together with the mechanical evolution step it sets the need for a new triangulation: the loop starts again until a stop condition is reached.
\item Triangulation. This step establishes the proximity relations among cells and is essential to reduce the computational complexity to ~ O(N).
\item Diffusion. The Delaunay triangulation performed in the previous step is used as a backbone for diffusion; the dual Voronoi construct is also needed. This step is discussed at length in the following sections.
\item Mechanical evolution. The cell cluster is held together by cellular adhesion forces and is subject to mechanical stress due to volume changes of individual cells as they grow and proliferate, or just shrink and dissolve after necrosis or apoptosis.
\end{itemize}

Here we see that the evolution of the cell cluster as a whole is indeed a very complex nonlinear process: growth is influenced by diffusion, growth changes the cluster structure and in turn this changes the way chemicals diffuse in the cluster. The different nutrients interact in the metabolic step and change absorption and consumption and this determines both cell growth and diffusion to neighboring cells \cite{cm2}. 
One additional detail -- included in the metabolic step --  conjures to make diffusion even more nonlinear than it might appear at first sight: while in most physical contexts one may safely assume the validity of Fick's law
\begin{equation}
\mathbf{J} = - D\nabla\rho
\end{equation}
where $\mathbf{J}$ is the diffusion current, $\rho$ is the concentration and $D$ is the diffusion coefficient, this is not true for the great majority of molecules in cell biology. Indeed diffusion is in most cases a complex process, mediated by transporters, specialized proteins that act as carriers across the cell membrane and the diffusion current is described by a Michaelis-Menten equation
\begin{equation}
\label{MMcurr}
J^{(out,in)} = \frac{J_{max}\rho^{(in,out)}}{K_m+\rho^{(in,out)}}
\end{equation}
where $J_{out,in}$ is the diffusion current (either outward or inward),  $J_{max}$ is the maximum diffusion current, $K_m$ is a specific constant which depends on the transporter, and $\rho_{in,out}$ is the concentration (either inside or outside the cell); this means that diffusion can be described correctly only at cell level, and that averaged mathematical descriptions necessarily miss this important point. The Michaelis-Menten diffusion kinetics is equivalent to Fickian diffusion at low concentration difference, but it starkly deviates from the linear behavior as it saturates at large concentration differences.
One result of this kind of transport is that diffusion across a membrane is much slower than diffusion in the cytoplasm, and this also means that the concentration $\rho$ is nearly uniform inside the cell.

\section{Solution of the diffusion problem in the cell cluster}
\label{sol}

Diffusion in the cluster is usually much faster than the mechanical rearrangement of single cells, and for this reason the solution of the diffusion problem can be obtained by balancing the input and output fluxes in the cells. This balancing is the working principle of the relaxation method used to solve diffusion equations on a square or a cubic lattice: the difference here is that the balancing is performed on a disordered graph -- the Delaunay triangulation of the cells' centers -- rather than on a regular lattice. The Delaunay triangulation provides a list of neighbors -- a cell communicates only with other cells in contact with it -- while the faces of the Voronoi regions approximate the contact surfaces. Thus the total mass change of a given diffusing substance in the $n$-th cell in the time interval $\Delta t$ is 
\begin{equation}
\label{diffeq}
\Delta M_n = \sum_{\langle neighbors \rangle} \left( J^{in}_{j,n} - J^{out}_{n,j}\right) A_{n,j} \Delta t - \lambda_n \Delta t
\end{equation}
where index $j$ denotes the $j$-th neighbor, $J^{in}_{j,n}$ is the incoming current from the $j$-th neighbor, $J^{out}_{n,j}$ is the outgoing current to the $j$-th neighbor, $A_{n,j}$ is the area of the Voronoi face between cells $n$ and $j$, and $\lambda_n$ is the absorption rate of the given substance in the cell ($\lambda_n$ can be either positive if the substance is consumed in the cell, or negative, if it is produced).
When a tumor spheroid is grown {\it in vitro}, it lives in a nourishing medium where diffusion also takes place, carrying, e.g. oxygen, from the atmosphere into the aqueous medium and then into the cell cluster: this means that the environment must also be included, and that the interface with the atmosphere and with the container walls sets the boundary conditions. To solve the diffusion problem we divide the aqueous medium in fictitious cells that we call {\it environmental cells}, and include them in the triangulation. Eventually we have a homogenous structure (cells + environmental cells) for the solution of the diffusion problem; environmental cells lying on the convex hull of the structure have known values of the diffusing substance and provide the boundary conditions. When the boundary conditions are constant in time, a stationary solution of the diffusion problem is obtained requesting that  $\Delta M_n=0$ for all cells. 

\section{Complexity of the mechanical evolution problem}
\label{mech}

Any simulation that involves $N$ elements with pairwise interactions has an $O(N^2)$ time-complexity: this means that simulations with large $N$ are dramatically slowed down and that one must find a way to cope with this problem. Solutions vary according to the nature of the simulated system, e.g. the N-body calculations in astronomy often use a hierarchical scheme that leads to an almost linear time-complexity (actually $O(N \log N)$) \cite{bu}. In the simulation of cell interactions we notice that any mechanical interaction is short-ranged, and that only cells that actually touch each other actually exert a direct force. We approximate cells with simple spheres with different radii, and to define the proximity relations we construct the Delaunay triangulation of cell centers. Since each cell has on average 12 neighbors (as in a close-packed cubic lattice), there are on average only $12 N$ interacting pairs, and this means that the 3D calculation of the Delaunay triangulation (which is an $O(N \log N)$ operation) and the force calculation has on the whole an $O(N \log N)$ time-complexity. 

\section{2D tests}
\label{2D}

\begin{figure}
\begin{center}
\includegraphics[width = 3in]{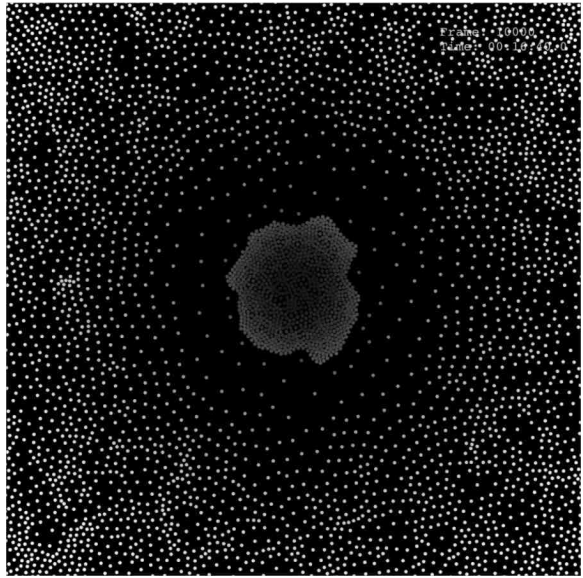}
\caption{Early phase in the calculation of mechanical relaxation and diffusion of a 2D system of cells. The cell cluster contains cells with a distribution of cell radii, and is surrounded by the environmental cells: in order to obtain a compact structure, cells are initially distributed at random in a large disk, and are surrounded by a similar distribution of environmental cells. We apply dummy forces that draw all the cells towards one another, and the configuration is allowed to relax. At the same time the diffusion is calculated as described in the text, and the gray levels represent different concentrations of a diffusing substance (here $O_2$; the lower the concentration, the darker the disk that represents a cell). When the structure has relaxed to a stable position we turn off the dummy forces and turn on the actual biophysical cell-cell interactions. In this example there are 1000 cells and 9000 environmental cells.}
\label{2DDiffusion-10000}
\end{center}
\end{figure}

\begin{figure}
\begin{center}
\includegraphics[width = 3in]{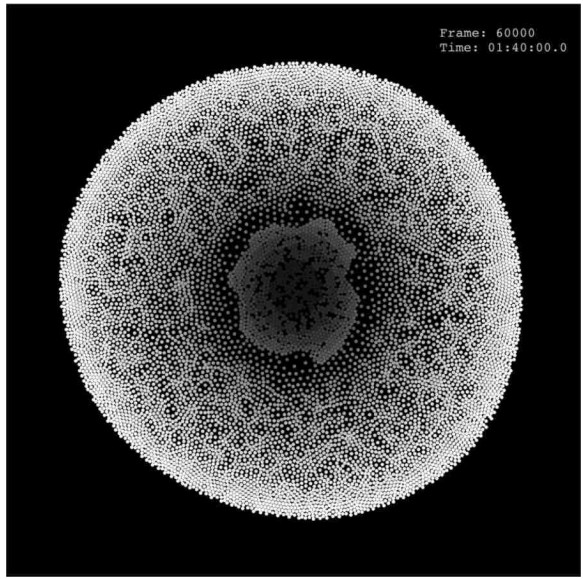}
\caption{Intermediate step in the calculation of the mechanical relaxation and diffusion. The figure represents a later stage in the evolution of the system of cells and environmental cells shown in figure \ref{2DDiffusion-10000}. Here it is apparent that the whole system has a roughly circular shape, and approximates a cell cluster floating in a drop of nourishing medium. The cells absorb oxygen and thus the oxygen concentration is lower inside the cluster, while it is maximum on the convex hull (the environmental cells in contact with atmospheric oxygen).}
\label{2DDiffusion-60000}
\end{center}
\end{figure}

\begin{figure}
\begin{center}
\includegraphics[width = 3in]{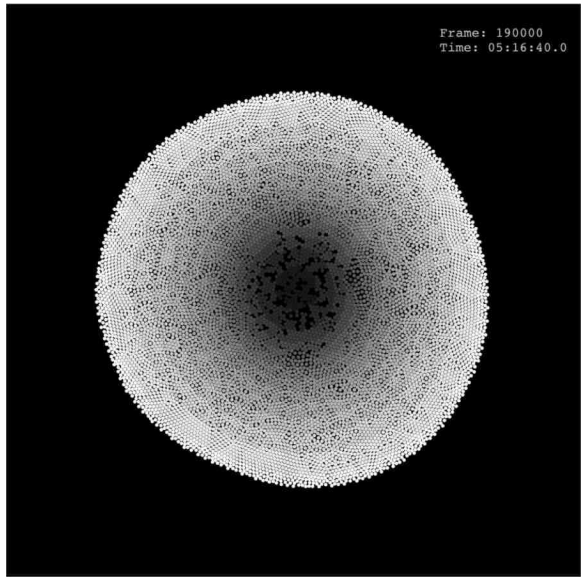}
\caption{This figure shows the relaxed configuration of the system of cells and environmental cells shown in figure \ref{2DDiffusion-10000}. In this simulation there is no proliferation, and thus no internal stresses develop and the cluster reaches a stable configuration.}
\label{2DDiffusion-190000}
\end{center}
\end{figure}

\begin{figure}
\begin{center}
\includegraphics[width = 3in]{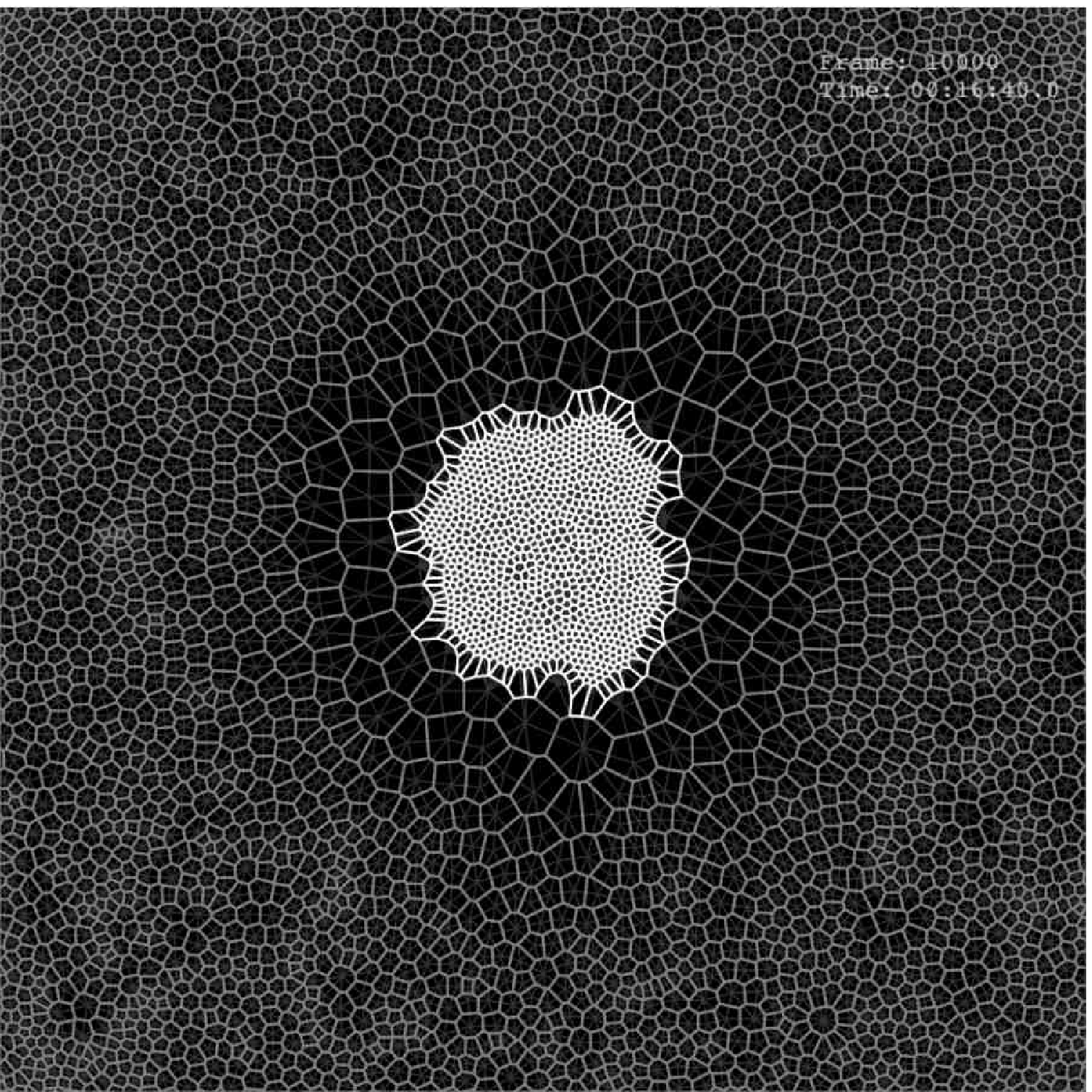}
\caption{This figure shows the Delaunay triangulation (background, dark gray), and the Voronoi tesselation (light gray for the regions that corresponds to cells in the cluster) that correspond to the distribution of cells in figure \ref{2DDiffusion-10000}.}
\label{QHTriangulation-10000}
\end{center}
\end{figure}

Comparison with experimental data requires a full 3D simulation that includes both the metabolic-driven Michaelis-Menten steps and pure Fickian diffusion on the Delaunay backbone, but  we have started with a prototype 2D version of the program \cite{cm} with a very simple approximation of metabolism and with Fickian diffusion only. The 2D version is written in C and uses Quickhull \cite{qh} for triangulations. Some results from this preliminary program are shown in figures \ref{2DDiffusion-10000}-\ref{2DDiffusion-190000} and \ref{QHTriangulation-10000}.

\section{3D implementation with CGAL}
\label{cgal}

The 3D program code which is presently under development is  written in C++; the cells in the simulation program are described by C++ objects and they include a set of internal state variables that define the cell state. Some of the internal state variables are associated to the metabolic state of the cell (this includes such parameters as the cell phase and the concentrations of important substances such as oxygen and lactic acid), while other variables describe the geometric and mechanical properties of the cell (such as cell center, cell volume, velocity of cell center, etc.).
The geometrical and mechanical variables depend on the Delaunay triangulation and the Voronoi tessellation, which are basic components of our simulator. As explained above, we need them both to define the proximity relations required by the dynamic evolution, and to carry out the calculation of diffusion in the cell cluster. The Voronoi tessellation  also yields a better approximation of the actual cell shapes, as a byproduct.

While several packages exist for 3D Delaunay and Voronoi calculations, not all are suitable for inclusion in our simulation program, because we need: 
\begin{itemize}
\item good computational efficiency;
\item robust implementation, free of precision issues;
\item easy access to the geometric structure;
\item dynamic update of existing triangulations;
\item C++ support;
\item good application support;
\end{itemize}

The CGAL library \cite{CGAL} fulfills all these requirements and is characterized  by ease of use, generality, efficiency and robustness.  The CGAL collaboration provides a well designed and maintained   library which also contains primitives, data structures and algorithms that are very useful in the development of  our code. In the near future, to describe the mechanical aspects of cell proliferation and death we shall need dynamic point insertion and deletion and CGAL is almost unique in providing insertion and deletion capabilities. 

If we return to equation (\ref{diffeq}) we see that the calculation of diffusion on the backbone provided by the Delaunay triangulation requires the explicit evaluation of areas of facets of the Voronoi cells and of their volumes (for the calculation of currents and densities), and in turn this needs a proper bookkeeping. 
A triangulation of a set of points in CGAL \cite{ct} is represented as a partition of the three-dimensional space into tetrahedra with four finite vertices (bounded tetrahedra) and fictitious tetrahedra (unbounded tetrahedra) having three finite vertices, forming a facet on the convex hull, whose fourth vertex is the fictitious infinite-vertex.
In the three-dimensional case the infinite-vertex forms infinite-cells with the facets of the convex-hull. The infinite-vertex has no coordinate representation and cannot be used for geometrical predicates. 
Delaunay triangulations, in three-dimensional space, are tetrahedralizations of points sets
which fulfill the empty sphere property stating that the circumsphere of any tetrahedron in the triangulation does not enclose any other vertex. Regular triangulations are defined for a set of weighted points where the weight corresponds to the square radius of the sphere associated to the point.  
In this data processing one looses the identity of the cells and hence, to obtain a mapping between the finite vertices and our cells, it is necessary to label  the vertices, for example using  the STL-container map. The data structure of CGAL returns the incidence relations between vertices, and thus it is possible, by means of an iterator, to find all the neighbors to a  finite vertex.
In this way we obtain the list of the  cells adjacent to a given cell. When the finite-vertex is adjacent to the infinite-vertex  the vertex is on the convex hull. One should have to pay attention to the degenerate cases in which different tetrahedrons have the same circumcenter, however we are interested in the non-degenerate cases of  random configurations of points.

\begin{figure}
\begin{center}
\includegraphics[width = 3in]{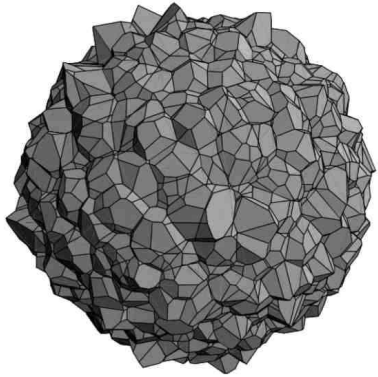}
\caption{This figure shows the result obtained in a test run with 2000 simulated cells in 3D (1000 cells + 1000 environmental cells). The Delaunay triangulation is calculated with CGAL, and we use the iterators and circulators provided by CGAL to construct the Voronoi tessellation. The open Voronoi regions, which correspond to cell centers on the convex hull of the cell clusters, have been removed because they are not useful for the calculation of diffusion: the cells on the convex hull are environmental cells and the concentration of oxygen and nutrients in these cells is a fixed function of time (this is the boundary condition for the discrete diffusion equation (\ref{diffeq})).}
\label{VoronoiBall}
\end{center}
\end{figure} 

CGAL provides for the three-dimensional triangulations iterators to visit all the cells, facets, edges and vertices and circulators necessary to iterate on the cells, facets and edges incident to a given vertex. Using these iterators and circulators we can to compute the dual diagram, i.e., the Voronoi tesselation associated to the Delaunay triangulation.
To construct the Voronoi diagram one may take the iterator over all the finite cells incident a given edge in the triangulation. Then, taking for every cell the dual, i.e. the center of  the circumsphere  of the four vertices,  we obtain all the faces in the Voronoi tesselation.
In this way, for every cell we find the set of neighboring cells, the distances of their centers from the cell center and the set of Voronoi points that define the cell boundary.
We can thus calculate the quantities needed by  equation (\ref{diffeq}), or, alternately by the Michaelis-Menten processes with transport equations like (\ref{MMcurr}).

Cells in the cluster are subject to pressure and adhesion to neighboring cells as well as to gravity, and they are also subject to random forces associated to Brownian motion and to fluctuations in adhesion forces. These forces are usually minute, but they are still able to bring about a slow mechanical relaxation of the cell positions, and they continuously reshape the cell cluster. For this reason neighboring time frames are expected to produce very similar triangulations and we plan to use the dynamical updating capabilities of CGAL to speed up the calculation of triangulations and tessellations.
Cells proliferate as well and the process of mitosis also contributes to the cluster rearrangement: mitosis also increases the number of cells, and when we include cell death (and finally cell dissolution) we see that the CGAL ability of dynamical center insertion and deletion may also speed up program operation and increase efficiency.

\section{Conclusion}
\label{conclusion}

The program for the simulation of tumor spheroids described in the previous sections has heavy and complex computational requirements. The inclusion of CGAL in the program framework has proved invaluable for the geometric part. The simulation program is quite complex and is still being developed, mainly in directions that do not involve computational geometry objects but rather the metabolic and proliferative description of cells \cite{cm2,thr}. In the future we shall profit greatly from some features that are unique to CGAL, like dynamic triangulation updates and point insertion and deletion capabilities.



\end{document}